\documentclass[11pt,a4paper,useAMS]{emulateapj}

\usepackage{natbib}
\usepackage{amsmath}
\usepackage{graphicx}
\usepackage{morefloats}

\def\nicefrac#1#2{
    \raise.5ex\hbox{#1}%
    \kern-.1em/\kern-.15em%
    \lower.25ex\hbox{#2}}

\begin{document}


\title{A distant radio mini-halo in the Phoenix galaxy cluster}


\author{R.~J.~van~Weeren\altaffilmark{1,$\star$},  H.~T.~Intema\altaffilmark{2},  D.~V.~Lal\altaffilmark{3}, F.~Andrade-Santos\altaffilmark{1},
M.~Br\"uggen\altaffilmark{4}, F.~de~Gasperin\altaffilmark{4}, W.~R.~Forman\altaffilmark{1}, M.~Hoeft\altaffilmark{6}, C.~Jones\altaffilmark{1},  S.~E.~Nuza\altaffilmark{7}, H.~J.~A.~R\"ottgering\altaffilmark{5}, and A.~Stroe\altaffilmark{5}\vspace{3mm}
}


\affil{\altaffilmark{1}Harvard-Smithsonian Center for Astrophysics, 60 Garden Street, Cambridge, MA 02138, USA}
\affil{\altaffilmark{2}National Radio Astronomy Observatory, 1003 Lopezville Road, Socorro, NM 87801-0387, USA}
\affil{\altaffilmark{3}National Centre for Radio Astrophysics, TIFR, Pune University Campus, Post Bag 3, Pune 411 007, India}
\affil{\altaffilmark{4}Hamburger Sternwarte, Gojenbergsweg 112, 21029 Hamburg, Germany}
\affil{\altaffilmark{5}Leiden Observatory, Leiden University, P.O. Box 9513, NL-2300 RA Leiden, The Netherlands}
\affil{\altaffilmark{6}Th\"uringer Landessternwarte Tautenburg, Sternwarte 5, 07778, Tautenburg, Germany}
\affil{\altaffilmark{7} Leibniz-Institut f\"ur Astrophysik Potsdam (AIP), An der Sternwarte 16, 14482 Potsdam, Germany}
\email{E-mail: rvanweeren@cfa.harvard.edu}

\altaffiltext{$\star$}{Einstein Fellow}

\shorttitle{Diffuse radio emission in SPT-CL J2344--4243}
\shortauthors{van Weeren et al.}

\vspace{0.5cm}
\begin{abstract}
\noindent 
We report the discovery of extended radio emission in the Phoenix cluster (SPT-CL J2344--4243, z=0.596) with the GMRT at 610~MHz. The diffuse emission extends over a region of at least 400--500~kpc and surrounds the central radio source of the  Brightest Cluster Galaxy, but does not appear to be directly associated with it. We classify the diffuse emission as a radio mini-halo, making it the currently most distant  mini-halo known. Radio mini-halos have been explained by synchrotron emitting particles re-accelerated via turbulence, possibly induced by gas sloshing generated from a minor merger event. {Chandra observations show a non-concentric X-ray surface brightness distribution, which is consistent with this sloshing interpretation}. The mini-halo has a flux density of $17\pm5$~mJy, resulting in a 1.4~GHz radio power of $(10.4\pm3.5)~\times 10^{24}$~W~Hz$^{-1}$. The combined cluster emission, which includes the central compact radio source, is also detected in a shallow GMRT~156~MHz observation and together with the 610~MHz data we compute a spectral index of $-0.84\pm0.12$ for the overall cluster radio emission. Given that mini-halos typically have steeper radio spectra than cluster radio galaxies, this spectral index should be taken as an upper limit for the mini-halo.

\vspace{4mm}
\end{abstract}



\section{Introduction}

Radio mini-halos are diffuse extended sources that are found in some relaxed, cool-core clusters  \citep[see the review by][]{2012A&ARv..20...54F}. They have steep radio spectra ($\alpha < -1$, with $S_\nu \propto \nu^\alpha$) and sizes of $\sim 100-500$~kpc, somewhat smaller than Mpc-scale giant radio halos that are found in merging galaxy clusters. The emission from mini-halos typically surrounds the central radio source associated with the Brightest Cluster Galaxy (BCG). Mini-halos are rare with less than 20 confirmed cases \citep[e.g.,][]{2012A&ARv..20...54F,2014ApJ...781....9G}. The prototypical example is the mini-halo found in the Perseus cluster \citep{sijbring_phd,2002A&A...386..456G}. 

Considering the lifetimes of the radio emitting electrons and the extent of the mini-halos, a form of in-situ cosmic ray (CR) production or re-acceleration is required to explain their presence. Several different possibilities have been put forward.  Hadronic (also called secondary) models invoke collisions between CR protons and thermal protons to produce a population of secondary relativistic electrons \citep{2004A&A...413...17P,2007ApJ...663L..61F,2010ApJ...722..737K,2013MNRAS.428..599F}. Alternatively, the synchrotron emitting electrons could arise from mildly relativistic electrons (for example cooled electrons from the central radio galaxy)  that have been re-accelerated by turbulence \citep{2002A&A...386..456G,2004A&A...417....1G}.

Recently, an interesting correspondence between cold fronts and mini-halos has been found for some clusters. Based on this, the idea has been put forward that sloshing motions, induced by minor mergers, generate turbulence in the cores of clusters, which is then capable of re-accelerating electrons. In the presence of a magnetic field, these electrons emit synchrotron emission and could form a mini-halo \citep{2008ApJ...675L...9M,2013ApJ...762...78Z}. {\cite{2014arXiv1401.7519B} also mention the possibility of a more direct connection and possible transition between mini-halos and giant radio halos in clusters.}

In this letter we report on the results of Giant Metrewave Radio Telescope (GMRT) radio observations of the Phoenix cluster (\object{SPT-CL J2344--4243}), which were taken as part of a larger radio survey of distant massive galaxy clusters \citep[i.e.,][]{2014ApJ...781L..32V}. 
The Phoenix cluster was discovered from its Sunyaev-Zel'dovich effect signal with the South Pole Telescope \citep{2011ApJ...738..139W} and is located at $z=0.596\pm0.002$ \citep{2012Natur.488..349M}.  The BCG appears to be undergoing a strong starburst with a star formation rate of $750\pm160$~M$_\sun$~yr$^{-1}$ \citep{2012Natur.488..349M}. This cluster is the most extreme cool-core cluster known \citep[$\dot{M} \approx 2700\pm700$ M$_\sun$~yr$^{-1}$;][]{2013ApJ...765L..37M} and is also very massive ($M_{500} \approx 1.3 \times 10^{15}$~M$_\sun$).  Interestingly, the combination of a high cooling rate and the relatively weak central radio source (associated with the BCG) suggests that feedback has been unable to halt cooling in this system, leading to the high star formation rate.

In this letter we adopt a $\Lambda$CDM cosmology with $H_{0} = 71$~km~s$^{-1}$~Mpc$^{-1}$, $\Omega_{m} = 0.27$, and $\Omega_{\Lambda} = 0.73$. With the adopted cosmology, 1\arcsec~corresponds to a physical scale of 6.652~kpc at $z=0.596$. {The resulting luminosity distance ($D_{\rm{L}}$) is $3495$~Mpc}. All our images are in the J2000 coordinate system.

\section{Observations \& data reduction}
\label{sec:obs}

SPT-CL~J2344--4243 was observed with the GMRT on June 14 and 15, 2013, for a total of  about 10~hrs. Due to the low-elevation of the cluster at the GMRT site, the observing time was divided into two separate sessions. The continuum observations were carried out at a frequency of 610~MHz with a usable bandwidth of 29~MHz. In addition, we used GMRT 156~MHz archival data taken on August 5, 2011, with a total on source time of  13~min. 
Both datasets were reduced via a semi-automatic data reduction scheme \citep{2009A&A...501.1185I} employing  the Astronomical Image Processing System\footnote{http://www.aips.nrao.edu} (AIPS), ParselTongue \citep{2006ASPC..351..497K} and Obit \citep{2008PASP..120..439C}. The data reduction started with  automatic radio frequency interference (RFI) removal, bandpass and gain calibration. The flux calibration was performed according to the scale described in \cite{2012MNRAS.423L..30S}. The calibration was further refined via several cycles of self-calibration. Low-level broadband RFI was subtracted using an improved version of the method described by \cite{2009ApJ...696..885A}. After self-calibration, direction-dependent calibration solutions were obtained for sources with a  signal to noise ratio $> 150-200$. The imaging was done via facets to apply the direction-dependent calibration solutions and to correct for the non-coplanar nature of the array.

Archival data from a \textit{Chandra X-ray Observatory} ACIS-I observation were also studied. The 11.9~ks of data  (\dataset [ADS/Sa.CXO#obs/13401] {ObsId 13401}) were calibrated following the processing described in \cite{2005ApJ...628..655V}, applying the most recent calibration files\footnote{We used CIAO 4.6 and CALDB 4.5.9}. The calibration included the application of gain maps to calibrate photon energies, a correction for position-dependent charge transfer inefficiency, and filtering of counts with a recomputed ASCA grade 1, 5, or~7 and those from bad pixels. In addition, periods of elevated background were filtered by examining the  light curves. Periods with count rates a factor of 1.2 above the mean count rate in the 6--12~keV  band were also removed. The discarded exposure time was negligible  (132~s). The final image was made in the 0.5--2.0~keV band  to increase the contrast between the thermal Bremsstrahlung and AGN emission from the central BCG, and to maximize the ratio between source and background counts. We used a pixel binning factor of 2.

\begin{figure*}[h]
\begin{center}
\includegraphics[angle =90, trim =0cm 0cm 0cm 0cm,width=0.47\textwidth]{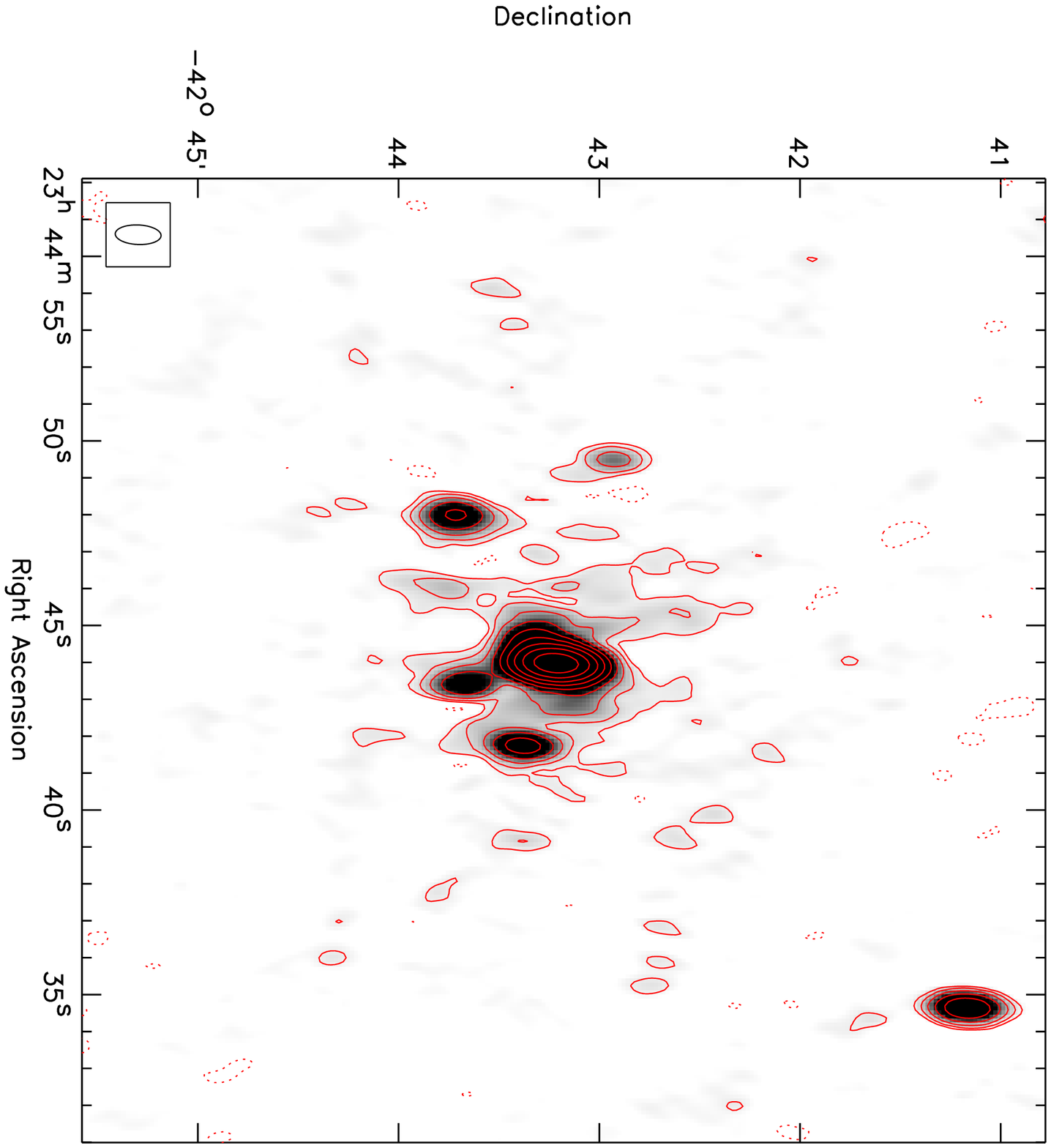}
\includegraphics[angle =90, trim =0cm 0cm 0cm 0cm,width=0.47\textwidth]{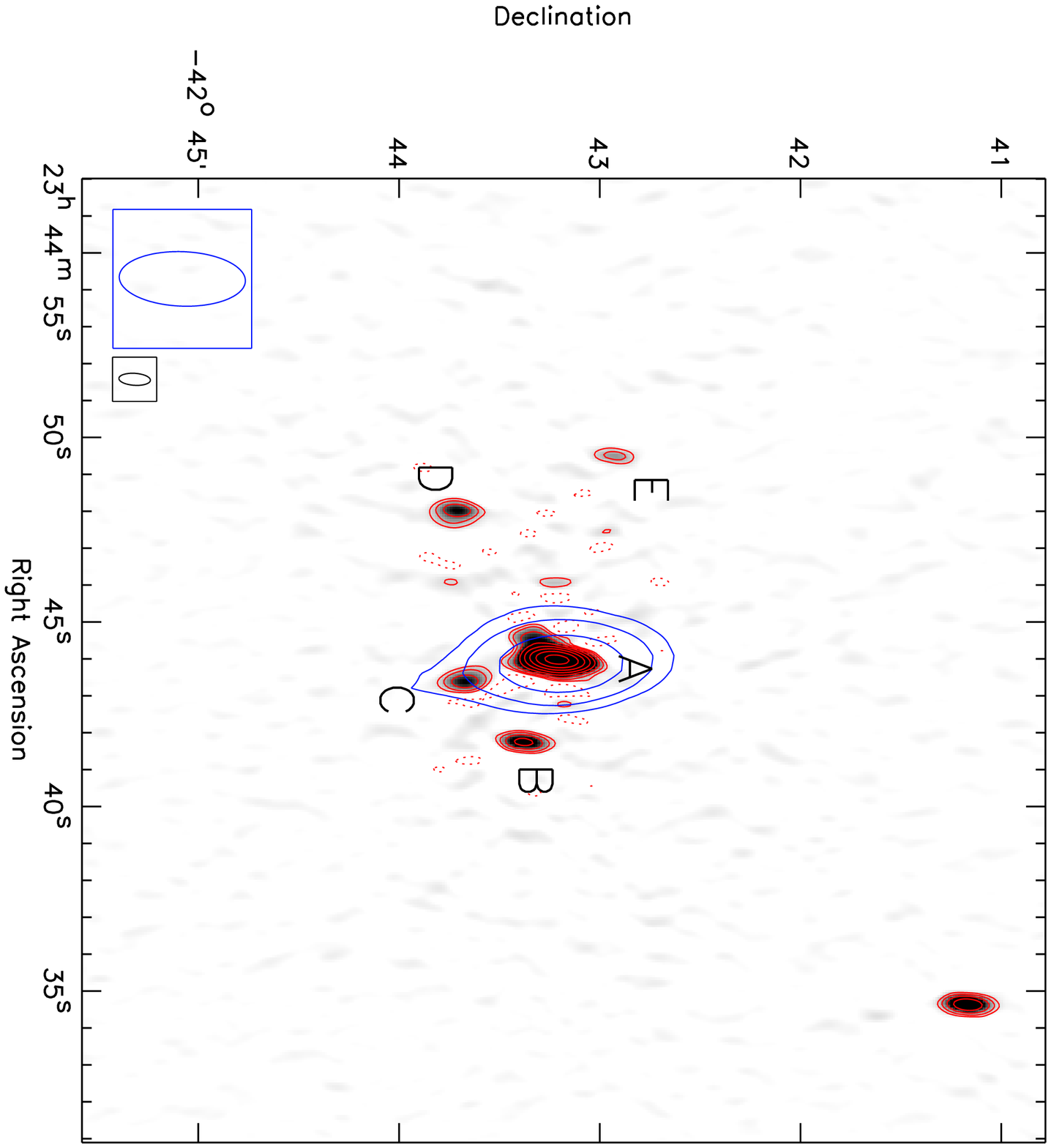}
\end{center}
\caption{Left: GMRT 610~MHz image of the Phoenix cluster. The image was made using Briggs weighing \citep{briggs_phd} with the robust parameter set to~0.5. Contour levels are drawn at ${[1, 2, 4, 8, \ldots]} \times 4\sigma_{\mathrm{rms}}$, with  $\sigma_{\mathrm{rms}}=43$~$\mu$Jy~beam$^{-1}$. Negative $-3\sigma_{\mathrm{rms}}$ contours are shown with dotted lines. The beam size is $13.7\arcsec \times 5.8\arcsec$ and indicated in the bottom left corner. 
Right: High-resolution ($9.4\arcsec \times 3.6\arcsec$) GMRT~610~MHz image is shown with red contours (and in grayscale). The contours are drawn at the same levels as in the left panel (but with $\sigma_{\mathrm{rms}}=95$~$\mu$Jy~beam$^{-1}$). {Compact sources are labelled}. For the imaging, data below 7~k$\lambda$ were excluded and the robust parameter was set to $0.0$. The GMRT 156~MHz image is shown with blue contours.  
The image has a central noise level of 8~mJy~beam$^{-1}$ and the contour levels are drawn at  ${[1, 2, 4, 8, \ldots]} \times 40$~mJy~beam$^{-1}$. The beam size is $38\arcsec \times 16\arcsec$.}
\label{fig:610im}
\end{figure*}

\begin{table}[t!]
\begin{center}
\caption{Cluster and radio mini-halo properties}
\begin{tabular}{ll}
\hline
\hline
$z$ &$0.596 \pm 0.002$\\
$R_{500}$ (Mpc) & 1.3\\
$L_{\rm{X, 500}}$ ($10^{44}$ erg~s$^{-1}$, 2.0--10.0~keV) & $82^{+1}_{-2}$ \\ 
$T_{500}$ (keV) & $13.0^{+2.5}_{-3.4}$ \\
$M_{500}$ ($10^{15}$  M$_{\odot}$ )& $1.26^{+2.0}_{-1.5}$ \\
$S_{\rm{mini-halo,610MHz}}$  (mJy) & $17\pm5$ \\
$S_{4\sigma_{\rm{rms}}~\rm{enclosed,610MHz}  }$  (mJy) & $114\pm11^{a}$ \\
$\alpha^{610}_{156}$ (compact and diffuse emission) &  $-0.84 \pm 0.12$    \\
$P_{\rm{mini-halo,1.4GHz}}$ ($10^{24}$ W Hz$^{-1}$)& $10.4 \pm  3.5^{b}$\\
LLS$_{\rm{mini-halo}}$ (Mpc) & 0.4--0.5 \\
610~MHz compact source fluxes (mJy) & $89.7^{c} \pm 9.0$ (A)  \\
& $4.29\pm0.44$ (B) \\
&   $3.46\pm0.36$ (C)\\
&  $3.95\pm0.41$ (D) \\
& $1.11\pm0.15$ (E) \\
\hline
\hline
\end{tabular}
\label{tab:properties}
\end{center}
Cluster properties taken from \cite{2012Natur.488..349M}\\
$^{a}$ {integrated flux density reported by the AIPS task {\tt TVSTAT} for all emission enclosed within the $4\sigma_{\rm{rms}}$ contours of Fig.~\ref{fig:610im} (left panel),  this includes sources A, B, C and the diffuse emission}\\
$^{b}$ assuming a spectral index of $\alpha=-1.1$ for the mini-halo emission\\
$^{c}$ source A has a peak flux of $70.5  \pm7.1$~mJy
\end{table}

\begin{figure*}[h]
\begin{center}
\includegraphics[angle =90, trim =0cm 0cm 0cm 0cm,width=0.47\textwidth]{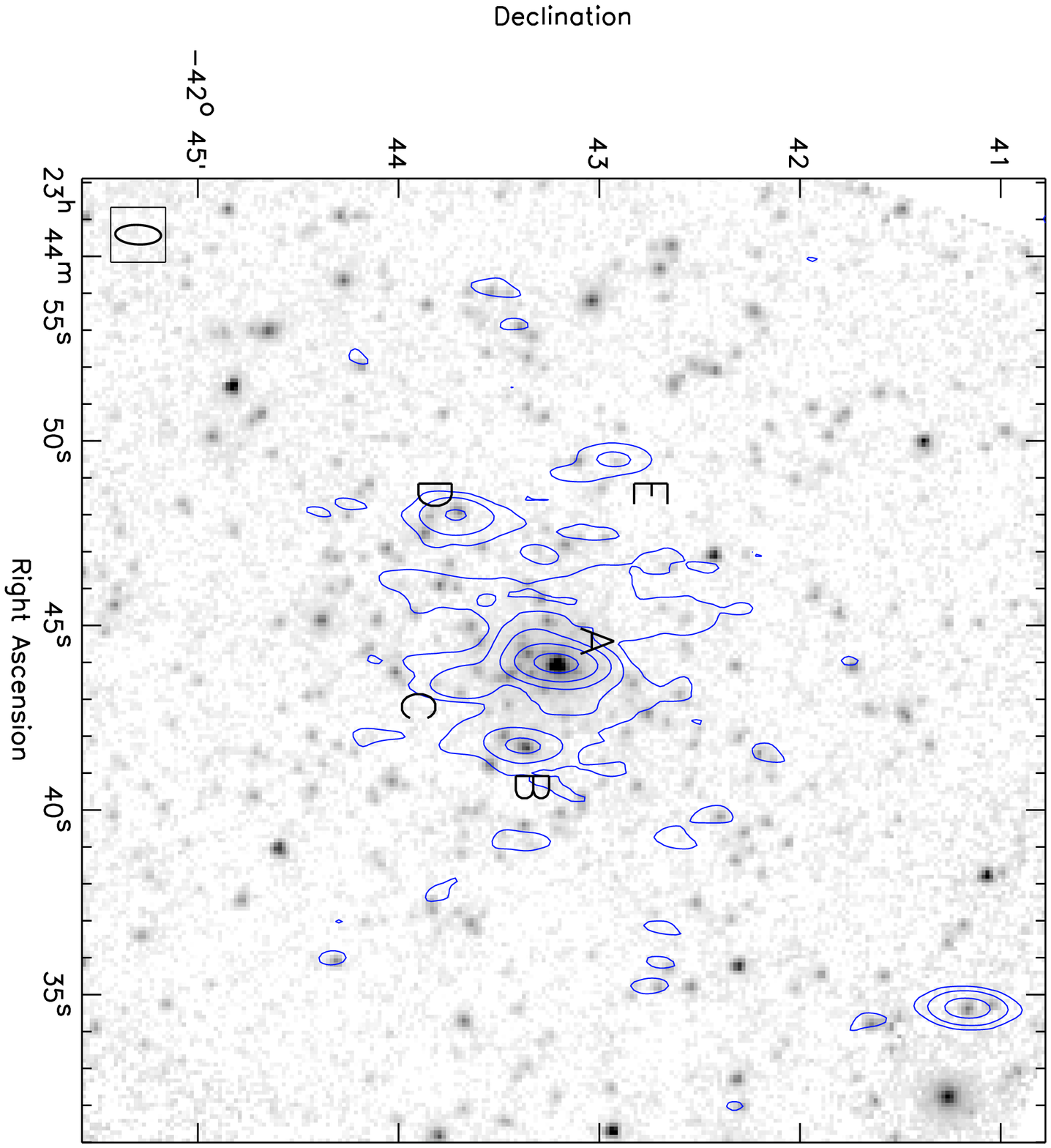}
\includegraphics[angle =90, trim =0cm 0cm 0cm 0cm,width=0.495\textwidth]{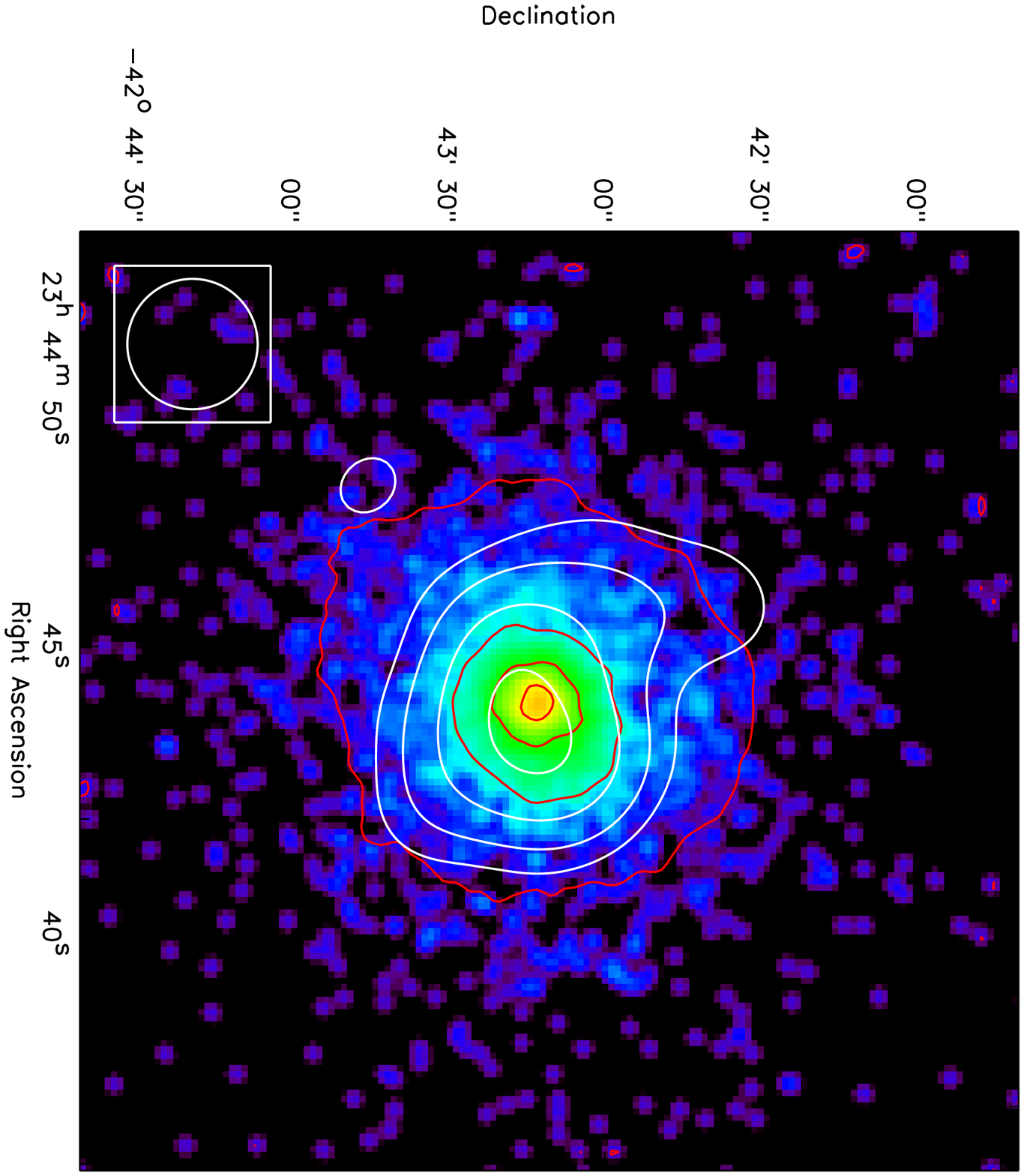}
\end{center}
\caption{Left: Spitzer 4.5~micron IRAC image overlaid with GMRT 610~MHz contours drawn at ${[1,  4, 16, \ldots]} \times 4\sigma_{\mathrm{rms}}$. Contours are from the image shown in Fig.~\ref{fig:610im} (left panel). Right: Chandra 0.5--2.0~keV ACIS-I image of the Phoenix cluster smoothed with a 3\arcsec~FWHM 
Gaussian (pixel size is 0.984\arcsec). {Red contours are X-ray isophotes drawn at levels of $[1,10,50,200] \times 10^{-5}$~cts~s$^{-1}$~pixel$^{-1}$.} GMRT 610~MHz low-resolution radio contours are overlaid in white. Compact sources (Fig.~\ref{fig:610im}; right panel) were subtracted from the uv-data prior to making this image. Contour levels are drawn at $\sqrt{[1, 2, 4, 8, \ldots]} \times 0.6$~mJy~beam$^{-1}$. The circular beam size is $25\arcsec$ and indicated in the bottom left corner. }
\label{fig:xray}
\end{figure*}

\section{Results}
\label{sec:results}
Our 610~MHz image of the Phoenix cluster, {with a resolution of $13.7\arcsec \times 5.8\arcsec$}, is shown in Fig.~\ref{fig:610im} (left panel). It reveals a central compact source associated with the BCG. This central source is also detected in the 843~MHz Sydney University Molonglo Sky Survey \citep[SUMSS,][]{1999AJ....117.1578B,2003MNRAS.342.1117M} with a flux density of $S_{843}= 79.2\pm2.8$~mJy. The 610~MHz image shows four additional sources within about 1\arcmin~of the BCG. {The image also reveals extended diffuse emission surrounding the central compact radio source. Subtracting the emission from compact sources and tapering down to 25\arcsec~resolution, we find a total extent of $\sim$0.8 to $\sim$1.2\arcmin~, with 1.2\arcmin~corresponding {to a largest linear size (LLS)} of 480~kpc. This low-resolution image is shown in  Fig.~\ref{fig:xray} (right), more details are given later in the Section.}  


The global cluster properties are summarized in Table~\ref{tab:properties}. A higher resolution image ($9.4\arcsec \times 3.6\arcsec$)  to emphasize compact sources was made by excluding baselines below 7~k$\lambda$ (corresponding to a spatial scale of about  0.6\arcmin) and is shown in the right panel of Fig.~\ref{fig:610im}. An overlay on an archival Spitzer 4.5~micron IRAC image is shown in Fig.~\ref{fig:xray} with compact radio sources labeled. It reveals counterparts to most of the radio sources visible in the high-resolution image. The GMRT 156~MHz image is overlaid in Fig.~\ref{fig:610im} (right panel). This shallow image only displays the central compact radio source, but there is a hint of a southward extension towards compact source~C.

Because of our limited spatial resolution and the fact that the cluster is located at $z=0.596$, it is somewhat difficult to accurately separate the diffuse emission from the central compact radio source.  In addition, the contribution from other compact sources needs to be subtracted. The integrated flux densities of the compact sources were measured with the AIPS task {\tt JMFIT}, fitting Gaussian components to the sources. The resulting integrated flux densities are given in Table~\ref{tab:properties}. The quoted uncertainties are based on the map noise and an absolute flux calibration uncertainty of 10\% \citep[e.g.,][]{2004ApJ...612..974C}.

The presence of diffuse emission on scales of $\sim 400-500$~kpc at 610~MHz, provides the first evidence that the central region of the Phoenix cluster contains an extended non-thermal component. In principle two possible explanations can be given for the diffuse component.  
The diffuse emission either originates from the radio lobes of the central AGN (source A) or it indicates the existence of a mini-halo. Since the diffuse emission surrounds the central AGN and its morphology does not resemble that of radio lobes, we classify this extended emission as a radio mini-halo. The physical extent of this mini-halo is also very similar to that of other known mini-halos in cool-core clusters \citep[e.g.,][]{2008A&A...486L..31C}. A small extension towards the SE of source A is seen in the high-resolution image (Fig.~\ref{fig:610im}; right panel). {We attribute this emission to source~A, and not to the mini-halo. Higher resolution radio observations are required to determine the precise nature of this extension.}

We estimate the integrated mini-halo flux density using two different methods. {For the first method, we simply  integrate the flux enclosed within the $4\sigma_{\mathrm{rms}}$ contours from Fig.~\ref{fig:610im} (left panel) and subtract the flux densities from the compact sources (A, B, and C) as are given in Table~\ref{tab:properties}. In this way we obtain a mini-halo flux density of $16.6 \pm 1.7$~mJy.} 
For the second method, we did attempt to remove the flux from the compact sources directly from the uv-data by subtracting the clean components from the high-resolution image shown in Fig~\ref{fig:610im} (right panel). We then re-imaged the data tapering to a resolution of 25\arcsec~resolution. This image is shown overlaid on the Chandra 0.5--2.0~keV X-ray image in Fig.~\ref{fig:xray}. The diffuse radio emission peaks approximately at the same position as the X-ray emission. From this low-resolution image we obtain a flux of $9\pm1$~mJy. This value likely underestimates the mini-halo flux density because some emission from the mini-halo has been subtracted since the mini-halo  contributes flux on scales larger than 7~k$\lambda$. If we double the inner uv-range cut for making the high-resolution clean component model, we obtain a flux of $19\pm2$~mJy for the radio mini-halo. 

{For the remainder of this Letter we adopt a flux density of $17\pm5$~mJy for the mini-halo. This was obtained by integrating within the $4\sigma_{\mathrm{rms}}$  contours and subtracting the compact sources flux densities}, but with the uncertainty of $5$~mJy reflecting the difficulties involved in separating the diffuse emission from the compact sources (i.e., approximately covering the range between the highest and lowest flux estimates reported in the previous paragraph). {With this integrated flux density, we calculate a $k$-corrected 1.4~GHz radio power of $ P_{\rm{1.4GHz}} = (10.4\pm3.5)~\times 10^{24}$ W~Hz$^{-1}$ using $ P_{\rm{1.4GHz}} ={4\pi S_{\rm{1.4~GHz}}D^{2}_{\rm{L}}} {\left( 1+z \right)}^{-\left(\alpha+1\right)}$. For the mini-halo spectral index we adopted a value of $\alpha = -1.1$ \citep[e.g.,][]{2014arXiv1403.2820G,2014ApJ...781....9G,2010A&A...509A..86M}}

At 156~MHz we measure a flux density $330\pm45$~mJy for the combined cluster emission (diffuse emission + compact sources), in a region twice the 156~MHz beam size centered on the peak emission. Taking precisely the same region at 610~MHz, we compute an overall spectral index (central AGN and mini-halo combined) of $-0.84 \pm 0.12$. Typically mini-halos have steeper radio spectra  than the central AGN and therefore the mini-halo emission could be considerably steeper than $-0.84$. In particular because the central compact core dominates the integrated flux density. However, resolved multi-frequency observations are needed to obtain separate spectral indices for the compact and diffuse components. In addition, deeper observations are needed to determine the full angular extent of the mini-halo.

{A closer inspection of the Chandra image reveals non-concentric X-ray isophotes (Fig~\ref{fig:xray}). The X-ray emission is more extended  in the NW direction at $\sim20\arcsec$ from the cluster center, while at a radius of $\sim1\arcmin$~the cluster emission extends more in the SE direction.}

\section{Discussion and conclusions}
\label{sec:conclusions}

We report the discovery of a $400-500$~kpc {(total extent)} radio mini-halo in the Phoenix cluster using GMRT 610~MHz observations. At a redshift of  $0.596 \pm 0.002$, the mini-halo in the Phoenix cluster is currently the most distant one known. Combining the  610~MHz data with shallow  156~MHz GMRT observations, we estimate an  upper limit of  $-0.84 \pm 0.12$ for the mini-halo spectral index. 
{By assuming a spectral index of $-1.1$ we estimated that the mini-halo has a  1.4~GHz radio} power of $(10.4\pm3.5)~\times 10^{24}$ W~Hz$^{-1}$,  higher than the Perseus cluster mini-halo \citep{sijbring_phd} but less than the RX~J1347.5$-$1145 mini-halo \citep[][]{2007A&A...470L..25G}. This radio power/luminosity falls in the range that is expected for mini-halos \citep[e.g.,][]{2014ApJ...781....9G}. 

\cite{2004A&A...417....1G,2007A&A...470L..25G} suggested a connection between the origin of the mini-halo synchrotron emission and cooling flows in clusters. This should result in a correlation between $P_{\rm{1.4GHz}}$ and the cooling flow power: $P_{\rm{CF}} =$ \nicefrac{$\dot{M}kT$}{$\mu m_{p}$}. {Such a correlation is observed \citep[e.g.,][]{2012ApJ...753...47D,2012AdAst2012E...6G}. For the Phoenix cluster we have $P_{\rm{CF}} = 3.5 \times 10^{45}$~erg~s$^{-1}$. Given this extremely high cooling flow power, we would expect  the Phoenix cluster mini-halo to have the highest known $P_{\rm{1.4GHz}}$ of any cluster, which is however not the case. However, it is important to note that the observed scatter is quite large for the correlation and also our derived cooling power is uncertain due to the uncertain mass inflow rate ($\dot{M} \approx 2700\pm700$ M$_\sun$~yr$^{-1}$). More accurate measurements of  $\dot{M}$ and also $P_{\rm{1.4GHz}}$ will be needed to determine whether the Phoenix cluster mini-halo deviates from the  $P_{\rm{1.4GHz}}$--$P_{\rm{CF}}$ correlation.}

The presence of a mini-halo in the Phoenix cluster has some interesting implications, related to the association of mini-halos, cold fronts and gas sloshing. Simulations show that cold fronts may be caused by gas sloshing, induced by a minor merger event \citep[e.g.,][]{2006ApJ...650..102A}. Minor mergers with a non-zero impact parameter result in cold fronts that form a non-concentric spiral-like pattern in the X-ray gas. These spiral-type patterns  have been observed in cool-core clusters 
and many of these clusters also host mini-halos \citep[e.g.,][]{2010ApJ...710.1776J, 2010A&A...511A..15L}.  In a few cases, the radio emission displays a spatial correlation with the spiral-like patterns in the X-ray gas \citep{2008ApJ...675L...9M}.  This indicates that the synchrotron emitting electrons are trapped in the same ICM flows that resulted in the spiral-like pattern. Since minor merger events also inject turbulence into cluster cores, it has been proposed that this turbulence could  re-accelerate a population of fossil electrons (for example from the  central radio galaxy) giving rise to diffuse radio emission in the form of a radio mini-halo \citep[see the numerical simulations by][]{2013ApJ...762...78Z}.

In the above framework, the presence of a mini-halo in the Phoenix cluster implies that the cluster core has enough turbulent energy, possibly caused by sloshing gas motions, to re-accelerate a population of relativistic particles. {This interpretation is also consistent with the non-concentric X-ray isophotes we found.} Since the gas-sloshing motions could be induced by a merger event, this raises the question whether the very high star formation rate measured for the Phoenix Cluster core could (partly) be the result of a merger event with a gas-rich system. This possibility is however  disfavored by \cite{2013ApJ...765L..37M}. Upcoming Chandra observations might have the sensitivity to detect cold fronts and spiral-like patterns to determine if the mini-halo in the Phoenix could be related to sloshing motions and turbulence.

\acknowledgments
{\it Acknowledgments:}
We would like to thank the anonymous referee for useful comments and the staff of the GMRT who have made these observations possible. The GMRT is run by the National Centre for Radio Astrophysics of the Tata Institute of Fundamental Research. This work is based in part on observations made with the Spitzer Space Telescope, which is operated by the Jet Propulsion Laboratory, California Institute of Technology under a contract with NASA.

R.J.W. is supported by NASA through the Einstein Postdoctoral
grant number PF2-130104 awarded by the Chandra X-ray Center, which is
operated by the Smithsonian Astrophysical Observatory for NASA under
contract NAS8-03060. H.T.I acknowledges support from the National Radio Astronomy Observatory, which is a facility of the National Science Foundation operated under cooperative agreement by Associated Universities, Inc. M.B and M.H. acknowledge support by the research group FOR 1254 funded by the Deutsche Forschungsgemeinschaft: ``Magnetisation of interstellar and intergalactic media: the prospects of low-frequency radio observations''. S.E.N. is supported by the DFG  NU~332/2-1.  W.R.F., C.J., and F.A-S. acknowledge support from the Smithsonian Institution. We thank Annalisa Bonafede and Julie Hlavacek-Larrondo  for commenting on an earlier version of this letter.



{\it Facilities:} \facility{GMRT}, \facility{CXO}


\begin{thebibliography} 
\expandafter\ifx\csname natexlab\endcsname\relax\def\natexlab#1{#1}\fi

\bibitem[{{Ascasibar} \& {Markevitch}(2006)}]{2006ApJ...650..102A}
{Ascasibar}, Y., \& {Markevitch}, M. 2006, \apj, 650, 102

\bibitem[{{Athreya}(2009)}]{2009ApJ...696..885A}
{Athreya}, R. 2009, \apj, 696, 885

\bibitem[{{Bock} {et~al.}(1999){Bock}, {Large}, \&
  {Sadler}}]{1999AJ....117.1578B}
{Bock}, D.~C.-J., {Large}, M.~I., \& {Sadler}, E.~M. 1999, \aj, 117, 1578

\bibitem[{{Briggs}(1995)}]{briggs_phd}
{Briggs}, D.~S. 1995, PhD thesis, New Mexico Institute of Mining Technology,
  Socorro, New Mexico, USA

\bibitem[{{Brunetti} \& {Jones}(2014)}]{2014arXiv1401.7519B}
{Brunetti}, G., \& {Jones}, T.~W. 2014, ArXiv e-prints

\bibitem[{{Cassano} {et~al.}(2008){Cassano}, {Gitti}, \&
  {Brunetti}}]{2008A&A...486L..31C}
{Cassano}, R., {Gitti}, M., \& {Brunetti}, G. 2008, \aap, 486, L31

\bibitem[{{Chandra} {et~al.}(2004){Chandra}, {Ray}, \&
  {Bhatnagar}}]{2004ApJ...612..974C}
{Chandra}, P., {Ray}, A., \& {Bhatnagar}, S. 2004, \apj, 612, 974

\bibitem[{{Cotton}(2008)}]{2008PASP..120..439C}
{Cotton}, W.~D. 2008, \pasp, 120, 439

\bibitem[{{Doria} {et~al.}(2012){Doria}, {Gitti}, {Ettori}, {Brighenti},
  {Nulsen}, \& {McNamara}}]{2012ApJ...753...47D}
{Doria}, A., {Gitti}, M., {Ettori}, S., {et~al.} 2012, \apj, 753, 47

\bibitem[{{Feretti} {et~al.}(2012){Feretti}, {Giovannini}, {Govoni}, \&
  {Murgia}}]{2012A&ARv..20...54F}
{Feretti}, L., {Giovannini}, G., {Govoni}, F., \& {Murgia}, M. 2012, \aapr, 20,
  54

\bibitem[{{Fujita} {et~al.}(2007){Fujita}, {Kohri}, {Yamazaki}, \&
  {Kino}}]{2007ApJ...663L..61F}
{Fujita}, Y., {Kohri}, K., {Yamazaki}, R., \& {Kino}, M. 2007, \apjl, 663, L61

\bibitem[{{Fujita} \& {Ohira}(2013)}]{2013MNRAS.428..599F}
{Fujita}, Y., \& {Ohira}, Y. 2013, \mnras, 428, 599

\bibitem[{{Giacintucci} {et~al.}(2014{\natexlab{a}}){Giacintucci},
  {Markevitch}, {Brunetti}, {ZuHone}, {Venturi}, {Mazzotta}, \&
  {Bourdin}}]{2014arXiv1403.2820G}
{Giacintucci}, S., {Markevitch}, M., {Brunetti}, G., {et~al.}
  2014{\natexlab{a}}, ArXiv e-prints

\bibitem[{{Giacintucci} {et~al.}(2014{\natexlab{b}}){Giacintucci},
  {Markevitch}, {Venturi}, {Clarke}, {Cassano}, \&
  {Mazzotta}}]{2014ApJ...781....9G}
{Giacintucci}, S., {Markevitch}, M., {Venturi}, T., {et~al.}
  2014{\natexlab{b}}, \apj, 781, 9

\bibitem[{{Gitti} {et~al.}(2012){Gitti}, {Brighenti}, \&
  {McNamara}}]{2012AdAst2012E...6G}
{Gitti}, M., {Brighenti}, F., \& {McNamara}, B.~R. 2012, Advances in Astronomy,
  2012

\bibitem[{{Gitti} {et~al.}(2004){Gitti}, {Brunetti}, {Feretti}, \&
  {Setti}}]{2004A&A...417....1G}
{Gitti}, M., {Brunetti}, G., {Feretti}, L., \& {Setti}, G. 2004, \aap, 417, 1

\bibitem[{{Gitti} {et~al.}(2002){Gitti}, {Brunetti}, \&
  {Setti}}]{2002A&A...386..456G}
{Gitti}, M., {Brunetti}, G., \& {Setti}, G. 2002, \aap, 386, 456

\bibitem[{{Gitti} {et~al.}(2007){Gitti}, {Ferrari}, {Domainko}, {Feretti}, \&
  {Schindler}}]{2007A&A...470L..25G}
{Gitti}, M., {Ferrari}, C., {Domainko}, W., {Feretti}, L., \& {Schindler}, S.
  2007, \aap, 470, L25

\bibitem[{{Intema} {et~al.}(2009){Intema}, {van der Tol}, {Cotton}, {Cohen},
  {van Bemmel}, \& {R{\"o}ttgering}}]{2009A&A...501.1185I}
{Intema}, H.~T., {van der Tol}, S., {Cotton}, W.~D., {et~al.} 2009, \aap, 501,
  1185

\bibitem[{{Johnson} {et~al.}(2010){Johnson}, {Markevitch}, {Wegner}, {Jones},
  \& {Forman}}]{2010ApJ...710.1776J}
{Johnson}, R.~E., {Markevitch}, M., {Wegner}, G.~A., {Jones}, C., \& {Forman},
  W.~R. 2010, \apj, 710, 1776

\bibitem[{{Keshet} \& {Loeb}(2010)}]{2010ApJ...722..737K}
{Keshet}, U., \& {Loeb}, A. 2010, \apj, 722, 737

\bibitem[{{Kettenis} {et~al.}(2006){Kettenis}, {van Langevelde}, {Reynolds}, \&
  {Cotton}}]{2006ASPC..351..497K}
{Kettenis}, M., {van Langevelde}, H.~J., {Reynolds}, C., \& {Cotton}, B. 2006,
  in Astronomical Society of the Pacific Conference Series, Vol. 351,
  Astronomical Data Analysis Software and Systems XV, ed. C.~{Gabriel},
  C.~{Arviset}, D.~{Ponz}, \& S.~{Enrique}, 497

\bibitem[{{Lagan{\'a}} {et~al.}(2010){Lagan{\'a}}, {Andrade-Santos}, \& {Lima
  Neto}}]{2010A&A...511A..15L}
{Lagan{\'a}}, T.~F., {Andrade-Santos}, F., \& {Lima Neto}, G.~B. 2010, \aap,
  511, A15

\bibitem[{{Mauch} {et~al.}(2003){Mauch}, {Murphy}, {Buttery}, {Curran},
  {Hunstead}, {Piestrzynski}, {Robertson}, \& {Sadler}}]{2003MNRAS.342.1117M}
{Mauch}, T., {Murphy}, T., {Buttery}, H.~J., {et~al.} 2003, \mnras, 342, 1117

\bibitem[{{Mazzotta} \& {Giacintucci}(2008)}]{2008ApJ...675L...9M}
{Mazzotta}, P., \& {Giacintucci}, S. 2008, \apjl, 675, L9

\bibitem[{{McDonald} {et~al.}(2013){McDonald}, {Benson}, {Veilleux}, {Bautz},
  \& {Reichardt}}]{2013ApJ...765L..37M}
{McDonald}, M., {Benson}, B., {Veilleux}, S., {Bautz}, M.~W., \& {Reichardt},
  C.~L. 2013, \apjl, 765, L37

\bibitem[{{McDonald} {et~al.}(2012){McDonald}, {Bayliss}, {Benson}, {Foley},
  {Ruel}, {Sullivan}, {Veilleux}, {Aird}, {Ashby}, {Bautz}, {Bazin}, {Bleem},
  {Brodwin}, {Carlstrom}, {Chang}, {Cho}, {Clocchiatti}, {Crawford}, {Crites},
  {de Haan}, {Desai}, {Dobbs}, {Dudley}, {Egami}, {Forman}, {Garmire},
  {George}, {Gladders}, {Gonzalez}, {Halverson}, {Harrington}, {High},
  {Holder}, {Holzapfel}, {Hoover}, {Hrubes}, {Jones}, {Joy}, {Keisler}, {Knox},
  {Lee}, {Leitch}, {Liu}, {Lueker}, {Luong-van}, {Mantz}, {Marrone}, {McMahon},
  {Mehl}, {Meyer}, {Miller}, {Mocanu}, {Mohr}, {Montroy}, {Murray}, {Natoli},
  {Padin}, {Plagge}, {Pryke}, {Rawle}, {Reichardt}, {Rest}, {Rex}, {Ruhl},
  {Saliwanchik}, {Saro}, {Sayre}, {Schaffer}, {Shaw}, {Shirokoff}, {Simcoe},
  {Song}, {Spieler}, {Stalder}, {Staniszewski}, {Stark}, {Story}, {Stubbs},
  {{\v S}uhada}, {van Engelen}, {Vanderlinde}, {Vieira}, {Vikhlinin},
  {Williamson}, {Zahn}, \& {Zenteno}}]{2012Natur.488..349M}
{McDonald}, M., {Bayliss}, M., {Benson}, B.~A., {et~al.} 2012, \nat, 488, 349

\bibitem[{{Murgia} {et~al.}(2010){Murgia}, {Govoni}, {Feretti}, \&
  {Giovannini}}]{2010A&A...509A..86M}
{Murgia}, M., {Govoni}, F., {Feretti}, L., \& {Giovannini}, G. 2010, \aap, 509,
  A86

\bibitem[{{Pfrommer} \& {En{\ss}lin}(2004)}]{2004A&A...413...17P}
{Pfrommer}, C., \& {En{\ss}lin}, T.~A. 2004, \aap, 413, 17

\bibitem[{{Scaife} \& {Heald}(2012)}]{2012MNRAS.423L..30S}
{Scaife}, A.~M.~M., \& {Heald}, G.~H. 2012, \mnras, 423, L30

\bibitem[{{Sijbring}(1993)}]{sijbring_phd}
{Sijbring}, L.~G. 1993, PhD thesis, University of Groningen

\bibitem[{{van Weeren} {et~al.}(2014){van Weeren}, {Intema}, {Lal}, {Bonafede},
  {Jones}, {Forman}, {R{\"o}ttgering}, {Br{\"u}ggen}, {Stroe}, {Hoeft}, {Nuza},
  \& {de Gasperin}}]{2014ApJ...781L..32V}
{van Weeren}, R.~J., {Intema}, H.~T., {Lal}, D.~V., {et~al.} 2014, \apjl, 781,
  L32

\bibitem[{{Vikhlinin} {et~al.}(2005){Vikhlinin}, {Markevitch}, {Murray},
  {Jones}, {Forman}, \& {Van Speybroeck}}]{2005ApJ...628..655V}
{Vikhlinin}, A., {Markevitch}, M., {Murray}, S.~S., {et~al.} 2005, \apj, 628,
  655

\bibitem[{{Williamson} {et~al.}(2011){Williamson}, {Benson}, {High},
  {Vanderlinde}, {Ade}, {Aird}, {Andersson}, {Armstrong}, {Ashby}, {Bautz},
  {Bazin}, {Bertin}, {Bleem}, {Bonamente}, {Brodwin}, {Carlstrom}, {Chang},
  {Chapman}, {Clocchiatti}, {Crawford}, {Crites}, {de Haan}, {Desai}, {Dobbs},
  {Dudley}, {Fazio}, {Foley}, {Forman}, {Garmire}, {George}, {Gladders},
  {Gonzalez}, {Halverson}, {Holder}, {Holzapfel}, {Hoover}, {Hrubes}, {Jones},
  {Joy}, {Keisler}, {Knox}, {Lee}, {Leitch}, {Lueker}, {Luong-Van}, {Marrone},
  {McMahon}, {Mehl}, {Meyer}, {Mohr}, {Montroy}, {Murray}, {Padin}, {Plagge},
  {Pryke}, {Reichardt}, {Rest}, {Ruel}, {Ruhl}, {Saliwanchik}, {Saro},
  {Schaffer}, {Shaw}, {Shirokoff}, {Song}, {Spieler}, {Stalder}, {Stanford},
  {Staniszewski}, {Stark}, {Story}, {Stubbs}, {Vieira}, {Vikhlinin}, \&
  {Zenteno}}]{2011ApJ...738..139W}
{Williamson}, R., {Benson}, B.~A., {High}, F.~W., {et~al.} 2011, \apj, 738, 139

\bibitem[{{ZuHone} {et~al.}(2013){ZuHone}, {Markevitch}, {Brunetti}, \&
  {Giacintucci}}]{2013ApJ...762...78Z}
{ZuHone}, J.~A., {Markevitch}, M., {Brunetti}, G., \& {Giacintucci}, S. 2013,
  \apj, 762, 78

\end{thebibliography}
\end{document}